\documentstyle[12pt,a4]{article}
\author{Yu.~V.~Lipko\\
        Institute of Solar-Terrestrial Physics SD RAS,\\
        p.~o.~box~4026, Irkutsk, 664033, Russia\\
        fax: +7 3952 462557; e-mail:~lipko@iszf.irk.ru}

\title{INHOMOGENEOUS STRUCTURE OF THE HIGH-LATITUDE IONOSPHERE
AS OBSERVED AT NORILSK}
\date{}
\begin{document}
\sloppy
\maketitle
\begin{abstract}
In March and August/September 1995, February 1996, and in March-April
1998, observations of the inhomogeneous structure of the high-latitude
ionosphere were carried out at Norilsk (geomagnetic latitude and
longitude are $64.2^\circ$N and $160.4^\circ$E, and $L=5.2$).
Small-scale irregularities (with the lifetime of several seconds,and
the spatial scale less than 5-7 km), and medium-size wave
irregularities(with the period of 10-50 min, and the horizontal size
of tens and hundreds of kilometres) of the ionospheric F layer were
investigated under different geophysical conditions. A total of 300
hours of observations was recorded, including 250 reflections from the
F2 layer, and the other reflections from the sporadic E layer.

The diurnal variations of
inhomogeneous structure parameters in March and  April is obtained.
Dependence of some ionospheric irregularity parameters on  geomagnetic
activity is presented.
\end{abstract}

\section{Introduction}
\label{intr}
It is known, that there are simultaneously irregularities of an
electron concentration of different spatial and time scales in the
high-latitude ionosphere [1]. The structure, dynamics, generation
mechanisms of such irregularities are different. This paper is devoted
to examination of small-scale irregularities (SSI) and medium-scale
travelling ionospheric disturbances (MS TID) at the high-latitude
region of Siberia. SSI are irregularities of electron concentration
with spatial scale of 0.1-10 km and time scale of several seconds
[2]. The main mechanism of generation of small-scale structure are
various sorts of instability in ionospheric plasma, associated with
electric fields and currents, which are especially strong in
high-latitude region [3]. MS TIDs have period of 10-40 minutes, and
the horizontal size of tens and hundreds kilometres. They are
ionospheric manifestations of acoustic-gravity waves (AGW), generated
in the lower termosphere [4].

The observations of inhomogeneous structure of an ionosphere in high
latitudes are carried out widely. For these purposes modern and
expensive facilities are used [5, 6, 7, 8]. However structure and
dynamic of ionospheric irregularities are investigated insufficiently.
It is explained by complexity and variability high-latitude
ionosphere. Besides there are huge areas, where the regular
observations of inhomogeneous structure of the ionosphere aren't
carried out. Such area is practically all high-latitude region of
Siberia. In the sixties-seventies at Norilsk the regular examinations
of small-scale structure high-latitude ionosphere by method D1 [9, 10,
11, 12] were carried out. The average patterns of motions in … and
F-layers of the ionosphere for various seasons of year and for levels of
geomagnetic activity were obtained. The studies of MS TIDs in Norilsk
were not carried out earlier.

Norilsk integrated magnitic-ionospheric station (geomagnetic latitude
and longitude $64.2^\circ$N and $160.4^\circ$E, $L=5.2$) operated by
the Institute of solar-terrestrial physics SD RAS is the Siberian
region's unique high-latitude station admirably equipped with an
appropriate facility for investigating the fine structure of the
ionosphere.

The objective of this study was to obtain average parameters both
small-scale irregularities and medium-scale TIDs depending on season,
local time and geomagnetic activity.

\section{Experimental equipment and technique}
\label{Equipment}
The measuring facility includes a digital vertical-incidence sounding
pulsed ionosonde [13]. The facility allows to receive of a
radiosignal, reflex from an ionosphere, on three antennas. After
analogue processing on each of three antennas amplitude, SIN- and COS-
components of a radiosignal, reflex from an ionosphere, were obtained.

Amplitude time series were processed by both the Similar-Fading Method
(SFM) and Correlation Method (CM) [9, 11]. SFM is based on calculation
of time delays between similar changes of amplitude of the
radiosignal. The SFM allows to determinate velocity and direction of
the motion of irregularities in the ionosphere. The CM using
characteristic ellipse (ellipse of anisotropy) allows to estimate the
shape, size, orientation and motion of SSI. These methods are
justified physically enough and supported by direct measuring using
incoherent scatter method. For many years ones were applied on the
network of stations (including Norilsk) to determinate parameters of
small-scale irregularities. The following parameters of small-scale
irregularities were calculated: velocity $v_s$ and direction $\alpha_s$,
size of
major characteristic ellipse half-axes $a$, degree of anisotropy $ex$,
lifetime of SSI $Th$.

SIN- and COS- components of the radiosignal are handled by phase
method [14, 15, 16]. The amplitude and phase spectrums of a signal,
reflex from the ionosphere, are calculated. Then time and space
derivatives are determined. They are used for estimation of direction
$\alpha_m$ and velocity $v_m$ of the motion medium-scale TIDs. The method is
applied since beginning of 70 years. Its adequacy is confirmed by
model calculations and comparison with known results. At the
high-latitude region of Siberia the phase method was used for the
first time.

\section{Experimental results}
\label{Experiment}
\subsection{ Dependence irregularity parameters on season and local
time}
The observations inhomogeneous structure of the polar ionosphere were
carried out in Norilsk from March 18 to April 13, 1998. It was
recorded about 160 hours of F2-layer radio-reflection. Obtained data
was classed into two seasons. The division into seasons was made on
base of results observations at Norilsk in the seventies.

The first period - equinox (March) - is characterise by high level of
magnetic activity (average summary $Kp=20$) and by the most strong
gradient of pressure, which arises from the fact of space-nonuniform
heating of an ionosphere by solar radiation. This month there is a
reorganisation from winter system of ionospheric plasma circulation to
summer one. The influence of solar radiation on processes of
ionisation and dynamics in the high-latitude ionosphere amplifies. In
the winter the ionosphere above Norilsk practically is not irradiated
by the Sun, the basic contribution to ionisation produce auroral
sources. In the summer the solar energy light on the large areas of
the high-latitude termosphere uniformly [11].

April, according to paper [11], belongs to the second period - summer.
This month the magnetic activity considerably weakens, the role
auroral processes is reduced, the role of solar radiation amplifies.
The average summer Kp coefficient per day equal 15.

The observations were carried out when geophysical conditions were
satisfactory (presence of qualitative reflection from the ionosphere)
and there were organisation opportunities. The time intervals of
observation were from 15 minutes to 10 hours. The interval of
processing both for  phase method and for  method D1 equal 40 s. The
values and root-mean-square deviations (r{.}m{.}s{.}) of ionosheric
irregularity (SSI and MS TID) parameters were averaged over  hour.
The values of direction $\alpha$ were calculated as the most probable values
for a hour.

Our procedure differs from one used at Norilsk earlier [11]. In the
seventies the averaging $\alpha_s$ and $v_s$
was carried thrice: over one minute
(time of a stationarity of the high-latitude ionosphere); over
interval of observation (5-6 min), and over every hour for every
season. The high-latitude ionosphere is characterised by the high
variability, thus three time averaging could give considerable error
for obtained parameters.

For March it was processed 72 hour of data (6500 intervals of
processing), received by reflection radiosignal from a F2-layer of the
ionosphere, for April it was processed 86 hour of data (7700
intervals). The correlation of 50 \% data was above 0{.}5, so these data
were handled by SFM. Approximately half of these data there was
suitable for processing by method CM, since degree of anisotropy $ex$
was positive for these.

Fig{.}~1 presents the diurnal variations of medium-scale TID parameters
for March and April. The dashed vertical line on the figures marks the
moment of passage solar terminator at 100 km altitude, where the
generation of acoustic-gravity waves takes place. Fig{.}~1~a,~e display
change of average value and r{.}m{.}s{.} of variations $f_d$ with time. For
frequencies of sounding, which were used (from 3{.}2 MHz up to 4 MHz),
0{.}1 Hz variation $f_d$ is matched to 7-8 m/s vertical velocity of moving
of the ionospheric layer as whole, the positive value variation $f_d$
associates with ionospheric layer going downwards, negative value -
upwards. Both for March and for April it is evident, that the peak
velocity of lowering (raising) of a layer in the sunrise and in the
sundown reach 15-20 m/s, at noontime  the layer as whole does not go
almost, at night the velocity of  raising is considerably reduced.

The width of the Doppler spectrum $\delta f_d$ is present
in the Fig{.}~1~b,~f.
Generally it does not exceed value 0.5 Hz. Before sundown $\delta f_d$ is
incremented that probably is explain by the generation of ionospheric
different-scale irregularities during passage of solar terminator
[17].

The most probable travelling direction MS TIDs in March and April was
southward. It was maintained practically stationary values during all
observation time (Fig{.}~1~c,~g). It is well accorded with results
obtained at the other stations. The various sources MS TIDs are given
in different papers. The analysis our data has not allowed to allocate
any source of generation MS TIDs. It is possible, as it is specified
in paper [4], the generation occurs at centre of the auroral oval. The
average velocity of travel MS TIDs $v_m$ was 50 m/s (see Fig{.}~1~d,~h).
After sundown the considerable deviation of velocity values was
observed.

Both March and April variations of parameters MS TIDs have general
features. March parameters are characterised major values r{.}m{.}s{.}
and more abrupt changes. It is probably associated with greater
auroral activity in this period and with major influence of solar
terminator.

Fig{.}~2 presents the variations of parameters of small-scale
irregularities in March and April. Fig{.}~2~a,~e presents variations
of degree of anisotropy ex, Fig{.}~2~b,~f -- variations of "lifetime"
$Th$. These parameters calculated using ellipse of correlation with
positive $ex$. Direction $\alpha_s$ and velocity $v_s$
of travel were obtained by SFM.

The degree of anisotropy $ex$, as it is visible from Fig{.}~2~a,~e, is 50
\%, i.e. the irregularities extend in the ratio 2{:}1 between major and
small half-axes of the ellipse. There is not the preferred direction
of major half-axis, but as have shown our examinations, irregularities
move across direction of major half-axis. Assumption, that
irregularities extend along magnetic field and move across it, does
not confirm in our experiments. The sizes of irregularities
appreciated by the major axis $a$ of the correlation ellipse are 400 m
in the morning and at noonday of April, being lowered up to 200 m at
the night. In March the average value of $a$ is 400 m. Average
"lifetime" $Th$ of irregularities in March practically does not vary and
equal 2 s (see Fig{.}~2~b), in April $Th$ raises from 1 s in the morning
up to 5 s in the evening and then $Th$ again decreases up to 2 s
(Fig{.}~2~f).

Fig{.}~2~c,~g present the direction $v_s$ of travel of SSI in March and
April accordingly. The light circles mark directions obtained in the
seventies [10, 11, 12], black circles mark direction obtained in 1998.
These directions coincide for March. Two-vortex system with western
convective jet in evening and east jet in morning sectors obtained in
the seventies, confirm that the essential role is played drift of
ionisation due to electric fields created by a magnetosphere
convection. The drift of irregularities in the morning and at the
noonday occur in northward direction, it is associated with the
gradient of pressure of solar terminator. In the evening the influence
of east electrojet is increased and by reason of that the SSI move in
westward direction. Then $\alpha_s$ changes to the eastward direction.
Interpretation of the direction of ionospheric plasma drift in April
it is more difficult. Practically during all measured period SSI move
in eastward direction. This direction does not coincide with direction
obtained in the seventies.

The average velocities of SSI travel $v_s$ are 100-150 m/s
(in Fig{.}~2~d,~h they are designated by black circles).
It is much higher than
velocities obtained in the seventies (are designated by light
circles). The considerable differences of velocity values can be
caused by the following moments. First, at the analysis in the
seventies it was used thrice averaging, that gives reduction of
average velocity. Secondly, the observations in 1998 were carried out
during the phase of solar activity growth, that, apparently, has had
an effect on magnification of velocity values.

Thus, as a result of the investigation the diurnal variations of
inhomogeneous structure parameters in March and in April is obtained.

\subsection{Dependence of parameters of ionospheric irregularities on
geomagnetic activity}
Fig{.}~3 presents dependence of ionospheric irregularity parameters on
geomagnetic activity. Geomagnetic activity was estimated by using
one-minute values of the coefficient of auroral electrojet intensity
AE.

Average variation of $f_d$, as shown on Fig{.}~3~a, is zero, but
r{.}m{.}s{.} $f_d$ is incremented from 0{.}1 Hz under €… 50 nT to 0{.}25 Hz
under €E greater, than 350 nT. The influence of auroral activity on
$\delta f_d$ is well appreciable (see Fig{.}~3~b). There is a considerable
broadening of $f_d$ spectrum from 0{.}4 Hz under €… is smaller, than 50 nT,
up to 0{.}8 Hz under €E exceeding 350 nT. The velocity of MS TID travel
$v_m$ (Fig{.}~3~c) under magnification AE up to 350 nT is incremented almost
three times, from 25 m/s up to 70 m/s. Under €… exceeding 350 nT $v_m$
decreases.

"Lifetime" $Th$ SSI under the increase of a level auroral activity
changes insignificantly. As the €… coefficient increase up to 150 nT
ex and a decrease, at the further magnification €… there is an inverse
process -- the $ex$ and $a$ are incremented. The velocity of travel of SSI
$v_s$ grows from 110 m/s up to 190 m/s under magnification €… up to 350
nT ( Fig{.}~3~d).

Thus, it is obvious, that auroral activity influences not only
on structure and dynamics of small-scale irregularities, that was
noted earlier, but also on behaviour medium-scale TIDs.

\section{Conclusions}
\label{Conclusions}
The following results were obtained:
\begin{enumerate}
\item The propagation directions and velocities of small -scale
irregularities and medium-scale TIDs are different. Medium-scale TIDs
travel predominantly in a southward direction with velocities of
40-100 m/s. The prevailing direction of small-scale irregularities is
eastward and westward, and their velocities lie in the range of from
100 to 200 m/s.
\item For the Norilsk region, for each season we obtained a diurnal
variation of averaged parameters of ionospheric irregularities of the
frequency Doppler shift; of the width of the Doppler spectrum;
propagation directions and velocities of medium-scale TIDs; and of the
anisotropy and lifetime of small-scale irregularities.
\item A dependence of medium-scale TID parameters on geomagnetic activity
was obtained. Auroral activity has a significant effect on the
frequency Doppler shift, the width of the Doppler spectrum and on the
propagation velocity of medium-scale TIDs.
\item The conclusions drawn on the basis of the observations from the 1970s
that auroral activity influences the drift of small-scale ionospheric
irregularities, specifically the 1{.}5-2--fold increase of the velocity
during the substorm, were confirmed.
\end{enumerate}

\section{Acknowledgements}
\label{Acnow}

We are indebted to V.V.Klimenko and to all staff members of the
Norilsk station for assistance in conducting the experiment. Special
thanks are due to E.L. Afraimovich and G.A. Zherebtsov for useful
discussions of the results.

\end{document}